\documentclass[]{interact}

\usepackage{color}
\usepackage{amssymb}
\usepackage{amsmath}
\usepackage{amsthm}
\usepackage[cal=boondox]{mathalfa}
\usepackage{ragged2e}
\usepackage{graphicx}
\usepackage{caption}
\usepackage{float}
\usepackage[caption=false]{subfig}% Support for small, `sub' figures and tables
\usepackage[natbibapa,nodoi]{apacite}
\renewcommand\bibliographytypesize{\fontsize{10}{12}\selectfont}

\theoremstyle{plain}% Theorem-like structures provided by amsthm.sty

\theoremstyle{definition}

\theoremstyle{remark}

\usepackage{algorithm}
\usepackage[noend]{algpseudocode}

\newcommand{\eq}[1]{(\ref{eq:#1})}

\newcommand{\sect}[1]{Section~\ref{sec:#1}}

\font\goth=eufm10 at 11pt

\def\Field{{\hbox{\goth F}}}

\def\Real{{\mathbb R}}   % Real numbers
\def\Esp{{\mathbb E}}

\def\th{{\theta}}

\def\given {\,|\,}

\begin{document}

\title{Pandemic model with data-driven phase detection, a study using COVID-19 data.}
\author{Yuansan Liu\textsuperscript{a}, Saransh Srivastava\textsuperscript{a}, Zuo Huang\textsuperscript{a} and Felisa J. V\'azquez-Abad\textsuperscript{a}$^\ast$ \thanks{$\ast$Also Professor of Computer Science,  Hunter College and Graduate Center CUNY, New York.}\\
\affil{\textsuperscript{a} School of Computing and Information Systems, The University of Melbourne.}}

\maketitle

\begin{abstract} The recent COVID-19 pandemic has promoted vigorous scientific activity in an effort to understand, advice and control the pandemic. Data is now freely available at a staggering rate worldwide. Unfortunately, this unprecedented level of information contains a variety of data sources and formats, and the models do not always conform to the description of the data. Health officials have recognized the need for more accurate models that can adjust to sudden changes, such as produced by changes in behavior or social restrictions. In this work we formulate a model that fits a ``SIR''-type model concurrently with a statistical change detection test on the data. The result is a piece wise autonomous ordinary differential equation, whose parameters change at various points in time (automatically learned from the data). The main contributions of our model are: (a) providing interpretation of the parameters, (b) determining which parameters of the model are more important to produce changes in the spread of the disease, and  (c) using data-driven discovery of sudden changes in the evolution of the pandemic.  Together, these characteristics provide a new model that better describes the situation and thus, provides better quality of information for decision making.
\end{abstract}


\section{Introduction}

The novel coronavirus (SARS-CoV-2) associated disease, COVID-19, was declared a pandemic infection within 3 months of the first case reported due to its fast spread world-wide that brought nations into standstill and forced people to stay indoors and observe distancing \citep{Yell:2020}. Almost all nations at some point have imposed some kind of restrictions to avoid the spread of the virus. Since the first cases described in Wuhan in the Hubai province of The People's Republic of China in December 2019, the global scientific community have actively researched and has given us some answers. Scientific developments abound and today the mechanisms by which viruses infect individuals are well known, although research is under way to fully understand the immune responses and defense mechanisms. At the time of submission, vaccines were still being developed \citep{MDVA:2020}.

%Public Health has provided guidance on how to prevent further contagion by measures that degrade the viral particle such as (i) wiping it using soap or hand sanitizers containing alcohol,  and (ii) implementing social distancing and use of face masks. The world is still suffering from a pandemic COVID-19 infection without precedent and  {\color{blue} at the time of submission of this paper} scientists around the world were working in the development of safe and efficacious vaccines \citep{MDVA:2020}.

 In this paper we address the problem as an epidemiology study, rather than a biochemical one. We use mathematical models based on the SIR model of \citep{abbey1952examination, kermack1932contributions} to describe the evolution of the pandemic.  These are based on ODEs (ordinary differential equations) that describe an ``aggregate'' behavior of the spread of the pandemic. 

Central to our study is the use of data. We live in unprecedented times where data is available freely via easily accessible internet repositories. .
 The sheer amount of data has triggered many publications examining and interpreting the evolution of the pandemic. Being humanly impossible to process all this data, analyses based mostly on epidemiology models, statistics and machine learning are now also growing and providing policy recommendations for NPI policies (non-pharmaceutical intervention). However many of these analyses yield contradicting results and it may become even more confusing for the general public. On May 2020, both Dr Anthony Fauci and Dr Deborah Birx (White House Coronavirus Task Force) recognized that the models are no longer useful for meaningful predictions. One of the main reasons is bad quality of data being used and how it is interpreted in the models. As well, Professor Rossi points out that ``{\em the models suffered from the same thing: they weren't robust enough to predict within the episodic, fat-tail-risk events}'' \citep{TheHill:2020}. In this paper we formulate a mathematical model that concurrently fits parameters and discovers such episodic changes.

Why are the state of the art epidemiology models not suitable for this pandemic? We believe that unprecedented data availability and the subsequent frequent regulatory changes make current models inappropriate. In this paper we address this research gap with two specific foci of attention.

The first focus of this paper is to be critical about the use and interpretation of data. We will deal with the facts that actual numbers are often under-reported, that data formats differ for different countries, and that the reported data may in some cases contain bias either intentional or unintentional. The second focus is to deal with the presence of sudden (or episodic) changes automatically by creating a data-driven method for knowledge discovery.
SIR-based models are sensitive to changes in social behavior and mutations of the virus. The behavior of the individuals in the society can be controlled by policy, regulations, restrictions and education. Establishing which of the model parameters is most affected by policy can provide insight for policy makers and health care providers in order to monitor the pandemic. Because these policies are applied at various times in different countries, the parameters of the model should be re-evaluated. In this paper we introduce a novel methodology that uses a statistical learning method of change detection in conjunction with the usual parameter fitting for the ODE model in order to produce a mathematical model that will discover random and sudden changes and accordingly adjust the parameters of the (otherwise deterministic) model.

Our model can provide information from the actual evolution of pandemics and thus provide new knowledge related to social behavior and infection dynamics. Can our model explain the impact of regulations? Can it explain the onset of particular mutations of the virus that have a significant impact? We will be answering these questions in the sequel.

\sect{lit} presents an overview of current literature on the topic. \sect{SEIDR} presents our SIR-type ODE model. \sect{stats} presents our first results that carry out statistical fitting using data. \sect{cpd} introduces a change point detection for discovering the actual times when regimes changed for various geopolitical locations. \sect{dpd} finalizes studying the methodology for making predictions based on our techniques that use both ODE models and statistical fitting of available data.

\section{Literature Review}
\label{sec:lit}

  As recently as 2020, there has been a surge of research papers trying to predict the trend COVID-19 is going to take.  Most of the epidemiology studies are based on the ODE methodology of the SIR model type. Although based on the evolution of ODEs, it is often implemented discretizing time, typically in the one-day scale. In \citep{fairoza2020coronatracker} the authors use a simple SEIR model to forecast the outbreak trajectory at the early stage of the pandemic. \citep{Sarkar2020} developed a modified SEIR model, where they divided the ODE into more compartments to analyze the pandemic in India, they did not consider death number due to the short epidemic time. \citep{NDAIROU2020109846} studied the pandemic in Wuhan using a modified SEIR model which added super spread, hospitalized, and fatal class. \citep{Glass2020} used a two-stage SEIR model, by setting different parameter values for pre and post lockdown, to study the effectiveness of lockdown in several countries, in their model, the death number was calculated from confirmed cases with a fixed mortality rate. In the CRISP model \citep{herbrich2020crisp}, which is based on the SEIR model, the authors use a probabilistic graphical model to predict infection spread. They go on to develop a Monte Carlo EM algorithm to infer contact-channel infection spread. The EpiLM model of \citep{vineetha2020individual}, also based on SIR compartmental framework, the authors provide tools for simulation and inference for discrete-time individual-level models, and \citep{Simuseir2020} simulated the COVID-19 epidemic based on SEIR model.
All publications listed above use the ``SIR'' type model, well accepted as mathematical models of infectious diseases, sometimes referred to as ``compartmental models''. This kind of models were first introduced by John Graunt as early as the $17^{th}$ century \citep{graunt1939natural}. The 1920's saw the rise of compartmental models for a closed population having fixed number of people. The Reed-Frost \citep{abbey1952examination} and Kermack–McKendrick epidemic model \citep{kermack1932contributions} \citep{kermack1991contributions} \citep{kermack1933contributions} gained significant importance, both describing the relationship among susceptible, infectious and recovered people, this kind of models are able to show how different public interventions may affect the outcome of pandemic.% {\color{red} Felisa: Baby explanation of ODE}
Many hybrid models were then created based on this basic SIR model. In \citep{Jaron}, the authors propose an epidemic model where the total population is divided into four types, S (susceptible, those able to contract the disease), E (exposed, those who have been infected but are not yet infectious), I (infected, those capable of transmitting the disease), R (removed, those who have become immune, dead or recovered), it later has been named as SEIR model.

  Others \citep{lorch2020spatiotemporal} design a more fine-grained spatiotemporal predictions of the course of the disease in the population. They make use of the data obtained through the contact tracing technologies and use real data for prediction.

  The work in \citep{Green} derives a two-group-SEIR model to study the effects of condom usage on preventing the HIV spread, where the author split the whole population into two different groups, assuming only people in one of the group will take the precaution(use condom). This approach provides inspiration for our study.

The ODE models aim at describing the evolution of the {\em number} of individuals in each class, approximated by a continuous variable. This is believed to accurately describe ``large'' populations that change by small amounts.
We thank the reviewers for pointing out the existence of discrete models for the spread of the virus. The research in \citep{Stoseir2020} presents a discrete stochastic model using SEIR model to study the efficiency of different social policies in COVID-19 pandemic. They model the number of individuals at each time step by generating random variables with discrete distributions (Poisson, and Binomial), where the mean values are chosen to fit the values that the ODE would predict. In \citep{fva-Chem:2015} the ODE approach is used in a different population model that describes the number of molecules of various proteins in a given cell. In that paper we show the relationship between a stochastic model for discrete space (where the transitions follow a multi-dimensional birth and death process) and the corresponding (continuous space) ODE model. We provide a mathematical proof that the ODE is not necessarily accurate when the number in the population is ``large'', but rather that under some assumptions the ODE values correspond to the {\em expected} number of individuals in the population. Exploring the impact of this relationship for pandemic evolution is not within the scope of the current paper. Simulation models also consider discrete space (number of individuals), few are agent-based simulation models and some of them are based on Markov chains. For example in \citep{covid-MTA:2021} we use a continuous time Markov chain to describe the transmission of the virus in short time scales (seconds and minutes). These approaches are different from the one we pursue in this paper. Specifically we aim to fill the gap that existing methods are not addressing directly, namely the fact that parameters change frequently  due to changes in policy and decision making, which  affects societal behavior and leads to sudden changes in the dynamics of the pandemic.

%{\color{red} Address the position of our method to fill the research gap here.}

% @article{fairoza2020coronatracker,
%   title = {CoronaTracker: Worldwide COVID-19 Outbreak Data Analysis and Prediction},
%   author = {Fairoza Amira Binti Hamzah and Cher Han Lau and Hafeez Nazri and Dominic Vincent Ligot and Guanhua Lee and Cheng Liang Tan and Mohammad Khursani Bin Mohd Shaib and Ummi Hasanah Binti Zaidon and Adina Binti Abdullah and Ming Hong Chung and Chin Hwee Ong and Pei Ying Chew and Roland Emmanuel Salunga},
%   journal = {Bull World Health Organ},
%   year = {2020},
%   doi = {http://dx.doi.org/10.2471/BLT.20.255695}
% }

% @article{mandal2020seqir,
%   title = {A model based study on the dynamics of COVID-19: Prediction and control},
%   journal = {Chaos, Solitons & Fractals},
%   volume = {136},
%   pages = {109889},
%   year = {2020},
%   issn = {0960-0779},
%   doi = {https://doi.org/10.1016/j.chaos.2020.109889},
%   url = {https://www.sciencedirect.com/science/article/pii/S0960077920302897},
%   author = {Manotosh Mandal and Soovoojeet Jana and Swapan Kumar Nandi and Anupam Khatua and Sayani Adak and T.K. Kar}
% }

\section{SEIDR model}
\label{sec:SEIDR}

%\subsection{Original SEIR model}

%Comparing to the SIR model, the SEIR model adds an ``exposed'' group to represent people that have the virus but are asymptomatic (some of them may test positive for the SARS-CoV2 virus, others are not tested, but all of these people have the virus). Adding this group is helpful when modeling the evolution of pandemics from a virus that causes a significant incubation period, in other words, people might not be symptomatic in the first few days after they caught the virus, but they can infect others. As is already known, the COVID-19 has an estimated incubation period of at most 14 days, which makes an SEIR based model more suitable for our task.
The situation simulated in \citep{Green} is similar to the current COVID-19 situation as there are two groups of people during the pandemic, one of them follows the policies and take precautions, while another group behaves as usual. For the COVID-19 situation, our model also considers two groups: those who take measures to prevent contagion (PPE, hygiene, social distancing) in group 1, and those who don't in group 2, and we extracted death number from the removed population in order to study the trajectory of the number of deaths. Then the whole population $N(t)$ is divided into five types: $S_{1,2}$ (susceptible in two groups), $E_{1,2}$ (people in two groups with the SARS-CoV2 virus, who can transmit the virus but are asymptomatic), $I$ (people who have COVID-19 symptoms), $R$ (recovered people who have antibodies), $D$ (dead). Consequently we use the acronym SEIDR. Here,. we extracted death number from the removed population in order to study the trajectory of the number of deaths.

\begin{figure}[h]
    \centering
    \includegraphics[scale=0.4]{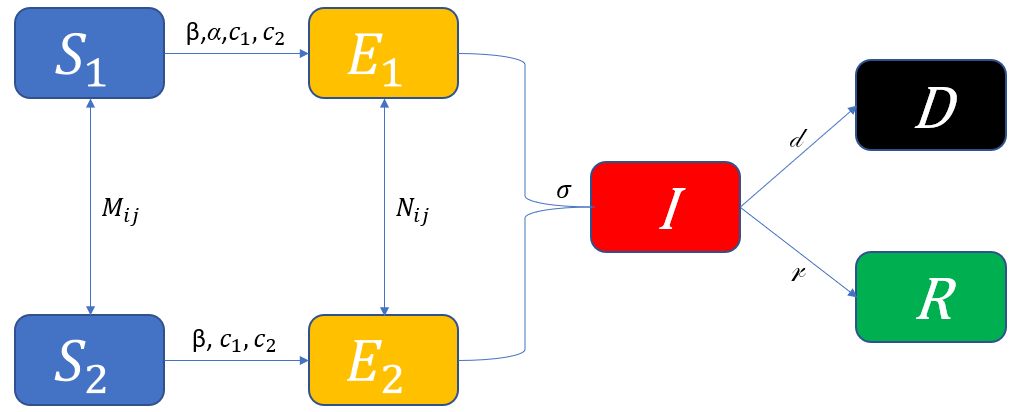}
    \caption{The flowchart of model}
    \label{fc}
   \end{figure}

Given a contact between a susceptible individual in Group 1 and another one that has the virus, the susceptible individual may or may not contract the virus. The parameter $\alpha \in [0, 1]$ models the lack of efficacy of preventive measures, as follows: (a) when following preventive measures, a susceptible individual (S) will be infected when in contact with the SARS-CoV2 with probability $\alpha$, and (b) an individual with SARS-CoV2 (E or I) will transmit the virus upon contact with another individual with probability $\alpha$, the case $\alpha = 1$ means that above preventive actions have no effect to prevent the virus. Thus, modeling random contacts in a population, the fraction of contacts between individuals in $S_1$ and $E_2$ that lead to infection is proportional to $(\alpha S_1) E_2$, whereas the fraction of contacts that lead to infection between individuals in $S_1$ and $E_1$ is proportional to $\alpha^2 S_1\, E_1$.
Figure \ref{fc} shows the structure of the model for a better understanding the dynamics. In this model, the parameters $c, M_{ij}, N_{ij}$ are all related to the government policy. Specifically, lockdown restrictions can decrease the parameters $c$, while forcing citizens to take precautions (mandatory use of masks, enforcing hygiene in public places) can increase $M_{21}, N_{21}$ which means more people getting into group 1. By changing these parameters, we can study the way that policy can affect the spread of the virus.

At the moment when we developed the model the general belief was that recovered individuals had antibodies and were immune to contracting the COVID-19 again. This is the reason why we used a feed-forward network model as shown in Figure \ref{fc}. The data that we used covers the initial period of the pandemic, where there was no evidence that recovered patients could contract the disease again, so this model is accurate for the data that we used. However, it is straightforward to adapt the model and introduce a loop back to the $S$ state after $R$.

In the model $E$ state always transitions into $I$. While it is true that individuals with the SARS-CoV2 virus may never acquire COVID-19 (the disease), the model reflects an aggregate flow, rather than describing each individual's path. Thus, when fitting our parameters for the network model, it will correctly reflect the data: a fraction of the transitions $E\to I\to R$ happen ``instantaneously''.

 We focused our initial efforts to reduce the dimension of the problem (fewer parameters) with the idea to simplify the fitting process, while retaining the essence of the pandemic spread. Based on BBC news, the number of people wearing masks increased during the pandemic\citep{bbc}. Thus, we assume people only migrate from the unsafe group to the safe one, but not in reverse direction. Second, we consider the pandemic dynamics during lockdown periods, when the population change due to migration is negligible. Hence, there are no changes in total population, and $M_{12} = N_{12} = 0$. The model for the spread of the virus is as follows:
 \begingroup
   \allowdisplaybreaks
   \begin{align}
    \frac{d S_1(t)}{dt} = & -\frac{\beta \alpha S_1(t)(c_1(\alpha E_1(t)+E_2(t))+c_2\alpha I(t))}{N(t)}+M_{21}S_2(t) \label{ss1}
\\
    \frac{dS_2(t)}{dt} =& -\frac{\beta S_2(t)(c_1(\alpha E_1(t)+E_2(t))+c_2\alpha I(t))}{N(t)}-M_{21}S_2(t)  \label{ss2}
\\
    \frac{dE_1(t)}{dt} =& \frac{\beta \alpha S_1(t)(c_1(\alpha E_1(t)+E_2(t))+c_2\alpha I(t))}{N(t)}%\nonumber\\&
    +N_{21}E_2(t)-\sigma E_1(t)\label{se1}
\\
    \frac{d E_2(t)}{dt} =& \frac{\beta S_2(t)(c_1(\alpha E_1(t)+E_2(t))+c_2\alpha I(t))}{N(t)}%\nonumber &\\
    -(N_{21}+\sigma)E_2(t) \label{se2}
\\
    \frac{d I(t)}{dt} =& \sigma(E_1(t)+E_2(t)) - (\mathcal{d}+ \mathcal{r}) I(t) \label{si}
\\
    \frac{d D(t)}{dt} =& \mathcal{d}I(t) \label{eq:sd}
\\
    \frac{d R(t)}{dt} =& \mathcal{r}I(t)\label{sr}\\
    P(t) =& \sum_{i=1}^{2} S_i(t) + E_i(t) + I(t) + D(t) + R(t) \label{tot}
\end{align}
 \endgroup
Where: \\
  $\beta$ : Contact rate between individuals of the population. \\
  $c_1$ : Average contacts with exposed individuals per person per day.\\
  $c_2$ : Average contacts with infected individuals per person per day. \\
  $\sigma$ : Incubation rate, at which infected people develop symptoms.\\
  $M_{ij}$ : Per capita rate of migration of susceptible people from group $i$ to $j$.\\
  $N_{ij}$ : Per capita rate of migration of infected ones from group $i$ to $j$.\\
  $\mathcal{d}$ : Death rate per day.\\
  $\mathcal{r}$ : Recovery rate per day.\\
  $\alpha$ : Probability of contagion or transmission given that restrictions are followed. \\

This model has $9$ parameters
\begin{equation}
\theta = (\beta, c_1,c_2,\sigma,M_{21},N_{21}, {\mathcal d}, {\mathcal r}, \alpha)
\label{eq:parameters}
\end{equation}

The {\em reproduction rate} $R_0$ of the virus is widely used in the literature. For the standard SIR models, this parameter is defined as $R_0=\beta/{\mathcal r}$ \citep{unravel}. Because our model also considers the rate of death from the COVID-19, we adapt the formula and use:
\[
R_0 = \frac{\beta}{{\mathcal d} + {\mathcal r}}.
\]

\section{Statistical Fitting}
\label{sec:stats}

\subsection{Selection criteria for datasets}
In  this section we focus on the experimental design that we carried out in order to validate our SEIDR model using available data. We use data from the COVID-19 dataset in \citep{actionsuser}, which source is maintained by Johns Hopkins University Center for Systems Science and Engineering (CSSE).

Fitting a model (in this case an ODE epidemiology model) using data presents important challenges. While it is true that an enormous amount of data is now publicly available all over the world, it is usually not in the same format, and it does not necessarily represent statistical samples from the variables in our model. For example,  the model describes the number of people at time $t$ that have been exposed to the virus. But there is no data for that. What we do have is the number {\em reported} to be infected every day. Clearly these are different variables, but dependent. Most of the COVID-19 available data records daily  (a) reported infected people, (b) reported dead people, and (c) reported recovered patients. \\

If we consider the data reporting process as a sampling from an unknown data stream, there are two factors that can affect the gap between the sampled data and the ``ground-truth'' data (model variables);  one is the sampling method, another is the volume of the sampling. In this study, we identify the following measures: (a) the number of tests per thousand people, which measures the capacity and method of the sampling process, and  (b) the reported infection numbers, which reflects the volume of the sample.

Validation of our model is done using data from a period of time to fit the parameters (see Section \ref{mf} below) and then using the model to simulate the evolution of the pandemic for some days after. These predictions are then compared with the actual data from the period after the one used in the fitting. It would be ideal at this stage  to use datasets from countries that report the number of infections, recoveries and deaths correctly, but there is no way to verify this, so we use the measures above in our selection criteria. Specifically, we chose countries where the number of tests per thousand people and the number of reported infections are the highest (as of June 09, 2020). For the period when we carried out this research. USA and UK were the top among such countries as shown in Table \ref{table:selection} \citep{owidcoronavirus}. We did not select Portugal data because it had less reported number of infections compared to the other three countries.
%\textcolor{red}{Although the model may not fit well in a dataset with low data integrity, we can still use it to indicate the sign of data inconsistencies.}

%{\color{cyan} If we consider the data reporting process as a sampling process from an unknown data stream, there are two factors can affect the gap between data samples and ground truth data, one is the sampling method, another is the volume of the sampling.
%
%In this study we seek to explore our model using data that are as much as possible a good representative of the model's variables. With this in mind, our data selection criteria look at the following: (a) number of tests per thousand people, and (b) reported infection number. The number of tests per thousand people reflects the capacity and the method of the sampling process, and the reported infection number reflects the volume of data sample. Based on the selection criteria above, we decided to choose countries where number of tests per thousand people and the reported infection are the highest (as of June 09, 2020)  \citep{owidcoronavirus}. USA and UK are among the top such countries as shown in table \ref{table:selection}.
%}\\

\begin{table}[h!]
\centering
\begin{tabular}{lrr}
\hline\hline
 Countries & \# of tests per 1000 & \# of reported infected \\ \hline
    \textbf{U.S.A.} &1.65&3,060,000\\
    \textbf{U.K.} &1.38&	288,977\\
    Portugal&1.29&44,859\\
%   \textbf{Italy}&0.47&242,149\\
\hline\hline
\end{tabular}
\caption{Number of tests per thousand people (rolling 7-day average) and the reported number of infections (cumulative) in 2020-06-09.}
\label{table:selection}
\end{table}

\subsection{Model fitting} \label{mf}
In order to fit the data, we used the cumulative number of deaths per day, because of the three most commonly reported numbers this is the one that better reflects the model's variable. Thus, as a working hypothesis we postulate that the available data is a sample that corresponds to the variable in the model. It is true that there may be some deaths of patients that were positive for COVID-19 but who also had pulmonary, cardiac and other severe conditions, so it is difficult to assess if the deaths were {\em caused} by the virus. On the other hand there might be deaths that were never reported as being caused by the virus. However these potential discrepancies are likely to be very small, so as a working hypothesis we assume that the daily number of deaths reported is an accurate number for the true daily number of deaths due to the virus.

Fitting the data using the number of deaths leads to solving an optimization problem for the parameter $\theta$ in \eq{parameters}. The mean square error (MSE) is given by:
\begin{equation}
J(\theta) = \frac{1}{T}\sum_{t=1}^{T} (\mathcal{D}_t - D_t(\theta))^2
\label{eq:optim}
\end{equation}
where $T$ is the number of data points used for training (in days), $D_t(\theta)$ is the cumulative number of deaths predicted at time $t$ (in days) by our SEIDR model for \eq{sd}, $\mathcal{D}_t$ is the corresponding available data for the cumulative number of deaths. The goal is to find the minimum of $J(\theta)$. %We implemented the algorithm in python leveraging its vast range of built-in packages. For integrating the system of ordinary differential equations, we use {\tt odeint} function from the {\tt scipy} package, and for curve fitting, we used the {\tt model} class from the {\tt lmfit} package that uses a deterministic non-linear least squares method known as LMA \citep{levenberg1944method}. A finite difference approximation is used to estimate the Jacobian of the function in LMA. The function {\tt odeint} solves a system of ordinary differential equations using {\tt lsoda} from the FORTRAN library {\tt odepack} which automatically switches between Adams and BDF methods depending on the behaviour of the problem.
Experiments show that our model has multiple local minima (due to non-convexity of the MSE function, sub-optimality and non uniqueness of the parameters). For comparison purposes this is not desirable, so we use constraints on the parameters to contain the solutions. Normally different constraints can lead to convergence of the model to different ``optimal'' parameters using the same dataset. However, the model can always produce consistent solutions if we set the constrains in reasonable ranges and the dataset is in good consistency like U.K. dataset. Table \ref{table:constrain of para} gives examples of the constrains of parameters that we used for fitting, but the readers can try other constrains to find a different fitting. Choosing appropriate constraints that lead to comparable models was done via experimentation and we will not include in this study the details of our procedure.

\begin{table}[h!]
\centering
\small
\begin{tabular}{lllll}
\hline \textbf{Parameters} & UK &US \\ \hline
    $\beta$ : Contact rate between individuals&$R_0(\mathcal{r}+\mathcal{d})$ &   $R_0(\mathcal{r}+\mathcal{d})$\\
    $R_0$ : Reproduction Rate &[2,5] &[2,5]\\
    $c_1$ : Average contacts per person per day&[0.5,2]& [0.1,2]\\
    $c_2$ : Average contacts per person per day&[0.1,0.8]&[0.01,0.6]\\
    $\sigma$ : Incubation rate &[0.005,0.3]& [0.001,1]\\
    $M_{21}$ : Migration rate of Suspected group 2 to 1 &[0.2,0.8]&[0.3,0.9]\\
    $N_{21}$ : Migration rate of Infected group 2 to 1 &[0.4,0.99] &[0.6,0.99]\\
    $\mathcal{d}$ : Death rate per day&[0.0006,0.02]&[0.0006,0.02]\\
    $\mathcal{r}$ : Recovery rate per day&[0.10,0.40]&[0.20,0.50]\\
    $\alpha$ : Prob. of protections not preventing  &[0.08,0.20]& [0.001,0.09]\\
\hline
\end{tabular}
\caption{Examples of constraints for the parameters.}
\label{table:constrain of para}
\end{table}

%\newpage
\subsection{Validation of the SEIDR Model for COVID-19 Data}
\begin{figure}[h!]
  \centering
  \includegraphics[scale=0.20]{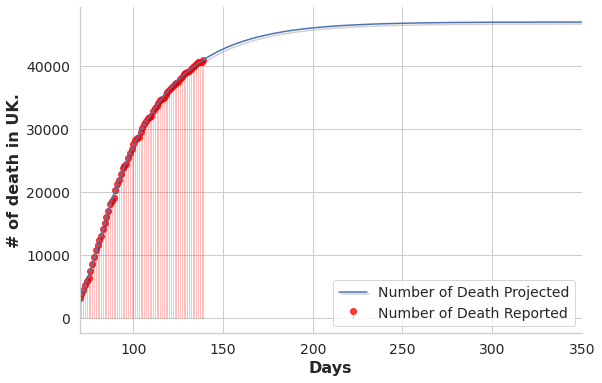}
  \includegraphics[scale=0.20]{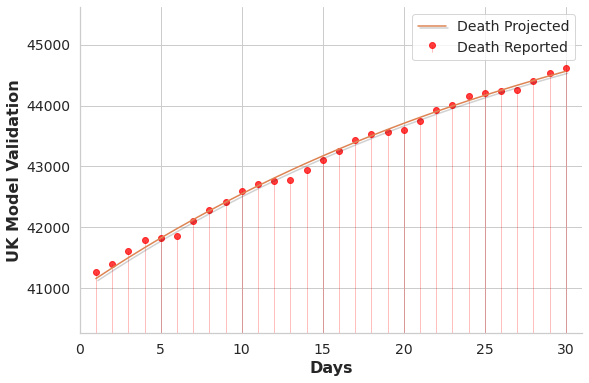}
  \includegraphics[scale=0.20]{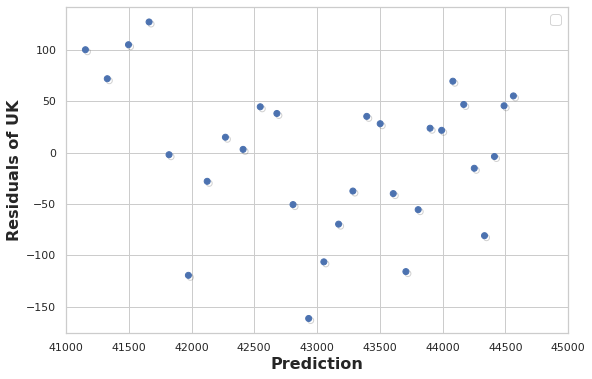}
  \caption{The left plot is the predicted number of deaths of UK model, red dots are reported number of deaths used to fit the model, from 2020-04-01 to 2020-06-09; middle plot is validation of the prediction of UK model in 30 days (from 2020-06-10 to 2020-07-09); right shows the residuals of the projection.}
  \label{Figure:uk}
\end{figure}

Our experimental design considers the data available for the U.K., from 1 April to 9 June 2020 (70 days)  to fit our model   parameter $\th$. After fitting the parameter $\th$, the data from 10 June to 9 July 2020 (30 days) is used to validate the model. Specifically, once the model parameter has been established, we used this value of $\th$ to run the ODE and predict the dynamics beyond the 10 June 2020 date. The resulting projections for the number of deaths are compared to the real data available for those days. Figure \ref{Figure:uk} shows the result of the validation, including the predictions (with our model) and the reported as well as the residuals.

\begin{figure}[h!]
  \centering
  \includegraphics[scale=0.20]{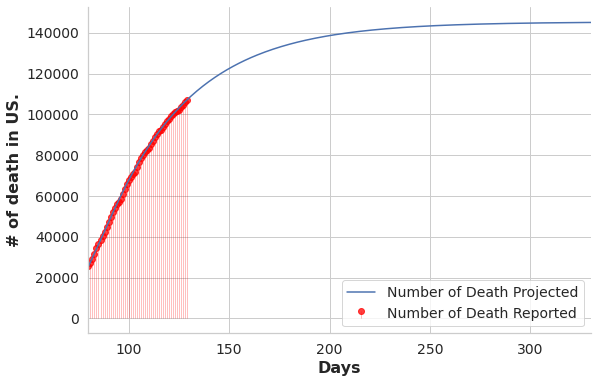}
  \includegraphics[scale=0.20]{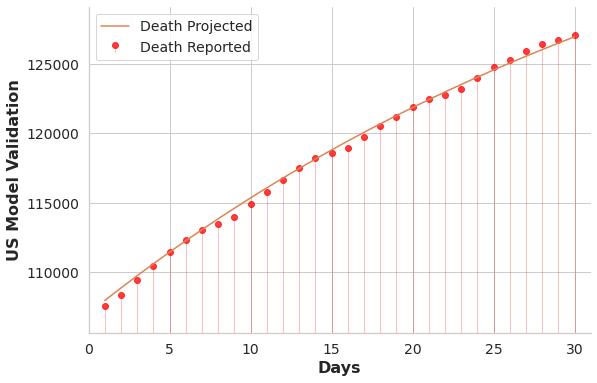}
  \includegraphics[scale=0.20]{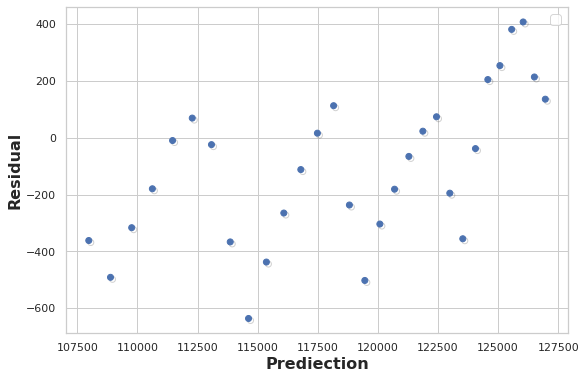}
  \caption{The left plot is the predicted number of deaths of US model from 2020-04-11 to 2020-05-30; middle is validation in 30 days (from 2020-5-31 to 2020-6-29); right shows the residuals of the prediction.}
  \label{Figure:us}
\end{figure}

Same design was also applied to U.S. data from 11 April to 20 June 2020, we use 50 days' data to fit the model. The corresponding plots are shown in Figure \ref{Figure:us}.
%\textcolor{red}{the data used to fit the model for the U.S. considers the period from 11 April to 30 May 2020 (50 days) and validation of this model shows a good fit using the predicted number of deaths against the data for the period from 31 May 2020 to 29 June 2020 (30 days).} .
% {\color{cyan}As to Italy, the data is from 22 March to 20 April 2020 and validation of this model shows a good fit as well with the data from 21 April 2020 to 20 May 2020 (30 days). The corresponding plots are shown in Figure \ref{Figure:ita}. }

% \begin{figure}[h!]
%   \centering
%   \includegraphics[scale=0.25]{IT1_d.png}
%   \includegraphics[scale=0.25]{IT1_val.png}
%   \includegraphics[scale=0.25]{IT1_res.png}
%   \caption{The left plot is the predicted number of deaths of Italy model from 2020-03-22 to 2020-04-20; middle is validation in 30 days (from 2020-4-21 to 2020-5-20); right shows the residuals of the prediction.}
%   \label{Figure:ita}
% \end{figure}

Considering our selection criteria we chose datasets of particularly good quality, so these results  strongly support the validity of our model for the COVID-19 pandemic. As mentioned before, it was not until later in the year that some recidivous cases were discovered. To describe this situation, our model can be modified for future studies, adding a transition from $R$ to $S$ to reflect the degradation of antibodies in some of the recovered patients, who lose immunity to the virus.

 \subsection{Vertical Adjustment}

\begin{figure}[h!]
  \centering
\includegraphics[scale=0.25]{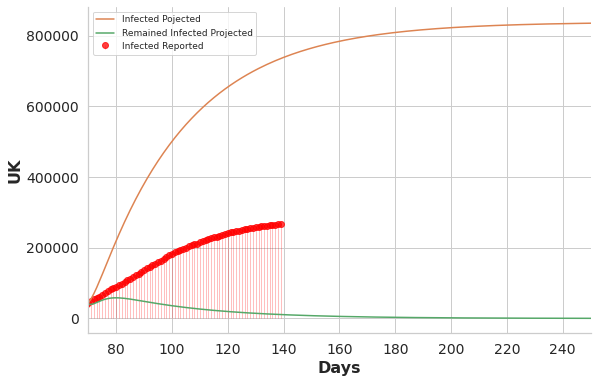}
\includegraphics[scale=0.25]{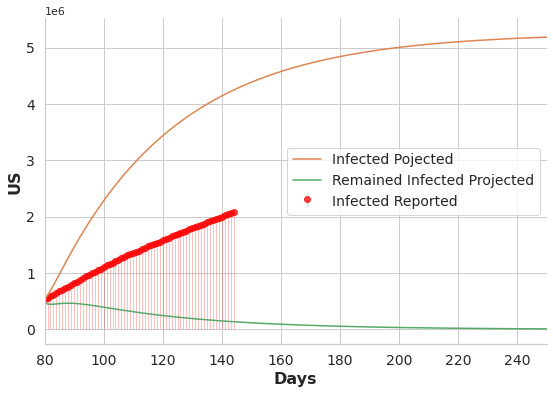}
\caption{Number of infections reported against data. Left is UK data from 2020-04-01; Right is USA data from 2020-5-31.}
  \label{Figure:v-gap}
\end{figure}

 Figure \ref{Figure:v-gap} shows the plots of the number of infected people predicted by the models against the available data of the number of {\em reported} infections. Clearly the reported number is much smaller than the number predicted by our model.

We believe this gap between the prediction and the data arises because (a) not everyone reports the infection, (b) not everyone is symptomatic and (c) not everyone is tested. Some people are not even aware of the infection and pass it in the community\citep{day2020covid} \citep{hu2020clinical}. In addition,  the reported infection number is also affected by the testing capacity of a country. \citep{owid} argues the importance of testing data, because more test cases will give us a better view of the spread.  From this rationale we can argue that if more people were tested, the reported number would increase to reflect more accurately the real situation, and it will also be closer to the predicted infection number from our model.

In this work, we propose a statistical correction for the model as follows. We use the vertical adjustment parameter $v$ (we will discuss in the following sections) to estimate the gaps between projected numbers and reported numbers. We use regression for the factor $v$, using reported data $\{r_t, t=1,\ldots, T\}$ and predicted data $\{p_t, t=1,\ldots T\}$ for the periods considered in the fitting and validation steps. The estimate is the result of  minimizing over $v\in\Real^+$ the mean squared error (MSE):
\[
\mbox{ MSE }(v) = \frac{1}{T} \sum_{t=1}^T (p_t-v\, r_t)^2.
\]

\begin{figure}[h!]
      \centering
      \includegraphics[scale=0.25]{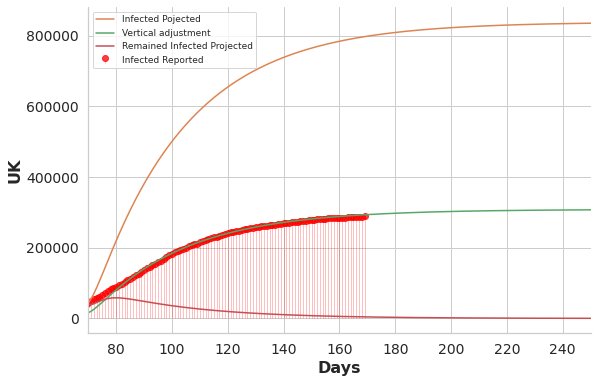}
      \includegraphics[scale=0.25]{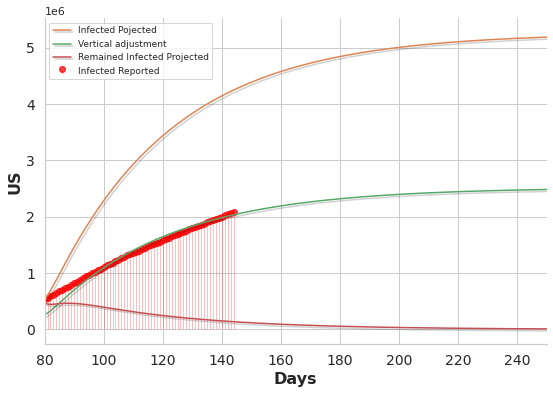}
      \caption{U.K., U.S. infection prediction with vertical adjustment}
      \label{va}
  \end{figure}

Figure \ref{va} shows the {\em vertical adjustment} of the model predictions, dividing the predicted infection numbers by the constant $v$. It is apparent  that the adjusted number is very close to the reported number, which justifies our belief that  there is significant  under reporting of infections, under the assumption that our ODE model is correct.

%\begin{figure}[h!]
%\centering
%  \includegraphics[scale=0.35]{UK_I.png}
%  \includegraphics[scale=0.35]{US1_I.png}
%  \caption{Left-hand side is the projection of UK model, red dots %are reported number of Infected, from 2020-04-01 to 2020-07-08. %Right side is US mode,from 2020-04-11 to 2020-08-1 }
%    \label{Figure:all_us_uk}
%\end{figure}\\

\subsection{Discussion}

Table \ref{table:parameter} shows the results of the model after fitting the parameters. The results of this analysis for various datasets can provide information to the specialists in charge of medical and social strategic management. Salient features that we can point out are  the similarity between various parameters of the model (that we also observed for other data). This is an indication that such parameters may depend highly on the biochemistry of the interaction between the virus SARS-CoV2 and the human respiratory system, or they may be correlated to the demographics and climate of the regions (USA and UK share many of these characteristics).  Notably
the parameters $\beta$, $c_1$, $M_{21}$, $\mathcal{r}$ and $v$ are similar in both countries.

It is interesting to notice that the parameter $R_0$ is about 3 times larger than the range of $0.7-1.0$  published by the Government Office for Science and Scientific Advisory Group for Emergencies of U.K. government \citep{rnumberUK2020}.

In the table we have highlighted the parameters that have significant  differences between countries.  Although we do not have the expertise in social and behavioral science, the differences do tell a coherent story. The death rate is larger in U.K., even when the average number of contacts is smaller. However the data shows a higher efficacy of preventive measures in USA (low $\alpha$) and a higher migration from infected individuals in Group 2 to Group 1 (thus the fraction of people following the preventive measures is higher in USA).

%\noindent{\bfseries Remark.} We have experimented extensively over the past four months.   Using data from Brazil, New Delhi and other locations we obtained similar results for the validation of our model. These plots and tables are not shown because of space limitation for the present contribution.

%{\color{cyan} The parameters fitted for the U.K. and USA data in table \ref{table:parameter} show a good level of independence {\color{red} ???}  to the models. The parameters $\beta$, $R_0$, $c_1$, $M_{21}$, $\mathcal{r}$ in are very similar in both countries. We noticed that the parameter $\mathcal{d}$ showed a strong dependence to the countries, the death rate in UK was much larger than US. The $R_0$ is about 3 times larger than the range of 0.7-1.0 published by the Government Office for Science and Scientific Advisory Group for Emergencies of U.K. government, (https://www.gov.uk/guidance/the-r-number-in-the-uk\#history)(to be added in reference). The $v$ parameter in \ref{table:parameter} shows that the number of infections is about 2.5 times larger than the reported number. }

\begin{table}[h!]
\centering
\begin{tabular}{lllll}
\hline \textbf{Parameters} & UK & US\\ \hline
    $\beta$ : Contact rate between individuals & 1.485& 1.027\\
    $R_0$ : Reproduction Rate & 4.966 & 4.030\\
    $c_1$ : Average contacts per person per day & 0.704 & 0.993\\
    $c_2$ : Average contacts per person per day& 0.124 & 0.199\\
    $\sigma$ : Incubation rate & 0.0517 & 0.0321\\
    $M_{21}$ : Migration rate of Suspected group 2 to 1 & 0.305&0.3000 \\
    $N_{21}$ : Migration rate of Infected group 2 to 1 & {\bf 0.439}&{\bf 0.989}\\
    $\mathcal{d}$ : Death rate per day & {\bf 0.01604}&{\bf 0.00483}\\
    $\mathcal{r}$ : Recovery rate per day & 0.283&0.250\\
    $\alpha$ : Prob. of protections not preventing  & {\bf 0.1999} & {\bf 0.0899}\\
    $v$ : Vertical Adjustment & 2.67&2.46\\
\hline
\end{tabular}
\caption{Model parameters of U.K. and U.S.}
\label{table:parameter}
\end{table}

\section{Phase detection}
\label{sec:cpd}

\subsection{Motivation}

\begin{figure}[h!]
  \centering
  \includegraphics[scale=0.30]{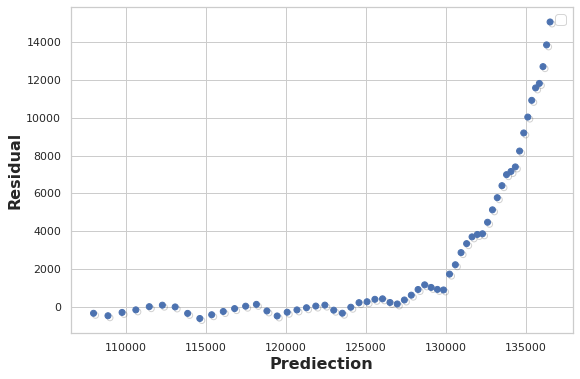}
  \includegraphics[scale=0.30]{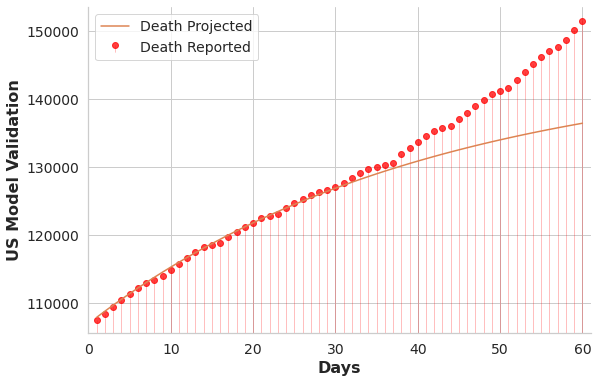}
  \caption{Model predictions and data for USA from 2020-04-11 to 2020-08-01. Left are the residuals and right are the projections.  }
  \label{Figure:us-all}
\end{figure}

We first present the results of the model versus the data for USA extending the horizon 30 more days than used for the validation period. To create Figure \ref{Figure:us-all} we use the parameter $\th$ that was obtained as the optimal value for the fitting problem that minimizes \eq{optim}, over the {\em training period} from 11 April to 30 May 2020. Using this value of $\th$, we calculated the model predictions beyond 30 May 2020, and compare with the available data.

While the model showed a very good fit for 30 day predictions (see Figure \ref{va}),  extending the analysis to 60 days clearly shows an anomaly. There is a significant increase in the amount of reported infections from day 150 (19-June) onwards, as shown in the plot to the right of Figure \ref{Figure:us-all}, and this leads to a huge divergence between the adjusted model prediction and the reported numbers. Looking at the residuals, where the fitting period was over the first 50 days, it is also apparent that after 06-July the parameters of the ODE model no longer describe accurately the evolution of the pandemic.

This pattern agrees with the fact that the epidemiology models of the SIR type contain parameters that depend on the social behavioral patterns, and on individual hygiene and prevention habits, in addition to the biological factors that govern the interaction between the human respiratory system and the SARS-CoV2 virus. In the 21st century the main patterns for the spread of pandemics are well understood and soon after the cases were first reported in the Chinese province of Wuhan, governments took measures all over the world (at different times) in order to contain the spread, or as it was commonly referred to, ``flatten the curve''.

In the particular case of the US data, to the naked eye infection rate changes occurred at the end of May, coinciding with Memorial Day, when many of the public areas usually open for festivities, including beaches, swimming pools, and outdoor cafes \citep{memday2020}.
The study by Center for Disease Control and Prevention \citep{salvatore2020recent} points out that increase in diagnostic tests cannot be the sole reason for observed increased cases. The reopening of universities and colleges for face-to-face classes possibly lead to transmission of virus among young individuals and their close contact. Also, the surge was more prominent in states where strict social restrictions were not imposed (like Texas) \citep{repState}.

But while the USA shows this surge in the Summer, other countries do not. Even within countries, data shows that imposition of social distancing and other measures have greater effect for some states or counties than for others. It is our view that instead of defining the phases according to published dates when policies have been put to effect, we will {\em let the data talk} in order to identify the changes in the phases of the pandemic.

The rest of this section introduces known statistical methods for change detection, and then we apply it to our model in order to automatically detect the phases of the ODE.

\subsection{Phase change detection on the COVID-19 data}
In this paper, we call a {\em phase} (for the ODE) a time interval where the model follows an ODE with constant parameters. It is apparent that sudden changes are imposed on the dynamics of the pandemic due to sudden changes in people's behavior. Notably restrictions including lockdowns, border closing, social distancing, self-isolation, obligatory use of masks and closing of certain businesses are placed in effect on specific dates in particular locations (countries or socio-political regions). Following the model, interpretable parameters such as $c_1,c_2, M_{21}, N_{21}$ should have different values on different phases. However, policy changes are not the only causes for shocks that can cause sudden changes in the process. Social behavior can also produce notable changes in the parameters, for example large numbers of people seeking unnecessary risk by voluntarily breaking the restrictions has been reported to cause abrupt surges in the number of people requiring medical care \citep{washPost:2020}.

Rather than defining the dates of the segments in advance and perform the statistical fitting of the SEIDR model within phases, we propose to ``let the data itself define the phases''. That is, we first apply change detection to the data in order to discover statistically significant points of changes in regimes (or phases). One of the potential benefits is assessing the effect of known shocks to the ODE, and discovering possible causes of changes that would not be easy to predict. In order to illustrate the method we now describe how we apply change detection to the data.

As mentioned before, the statistical fitting is done directly using the data on the number of deaths per day in order to find the parameters of the SEIDR model. It is on the {\em residuals} of the predicted cumulative number of dead people (by fitted ODE) that we will apply the change detection. Under the null hypothesis that the ODE has constant parameters throughout the whole sample points $(X_1,\ldots,X_n)$ the residuals (difference between the predicted data and the observed data) should have mean zero. Under this assumption we wish to define an appropriate CUSUM process for change point detection. The rationale is that if the true process has different phases, then the residuals would have mean different from zero on different phases (ill fitting). However we do not know in advance what the new means would be, even if the overall mean residual is zero (by construction, if the optimization procedure achieves a zero error in the approximation).

%The CUSUM method of \sect{CUSUM} requires knowledge of both the current mean and the mean for the next phase. Clearly this is not the case for our problem. While we can assume that the residuals have mean zero for the fitting period, we do not know if the following phase will be a surge or a decrease of cases.  For this reason,  we now  turn to one of the earlier descriptions of the point detection

We consider the method originally proposed by Page \citep{page1954continuous} and used for control charts \citep{montgomery2020introduction}. The algorithm defines in parallel two CUSUM processes: one to detect a positive mean and the other, a negative mean. The processes evolve as follows
\begin{align*}
  C_p(n+1) &= \max(0, C_p(n) + X_{n+1} - (\mu_0 + K))\\
  C_m(n+1) &= \max (0, C_m(n) - X_{n+1} + (\mu_0 - K)),
\end{align*}
where the starting values are $C_P(0)$ = $C_m(0) = 0$. The parameter $\mu_0$ is the original process mean and $K$ is called the reference value (or the slack value). Again the rejection region for the hypothesis is defined by a threshold $h$, and the stopping time
\[
T = \min\{ k \geq 0 \:  \colon\: \max(C_p(k), C_m(k))  \geq h\},
\]
is declared as the estimated change point. When the process that reaches the threshold is  $C_p$ the positive change is declared, otherwise the negative change is declared.
In \citep{montgomery2020introduction} the general control chart practice suggested is to choose $K = \frac{\mu_0}{2}$ and $h =  5\sigma$.

\begin{figure}[h!]
\centering
  \includegraphics[scale=0.3]{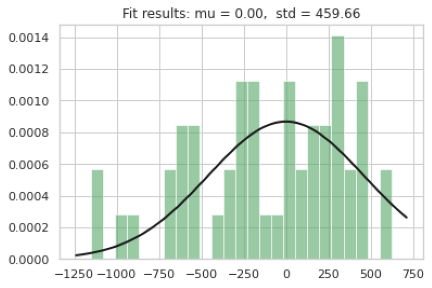}
  \includegraphics[scale=0.3]{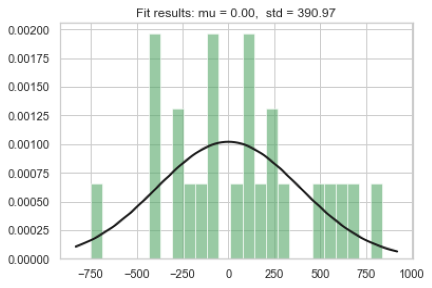}
  \includegraphics[scale=0.3]{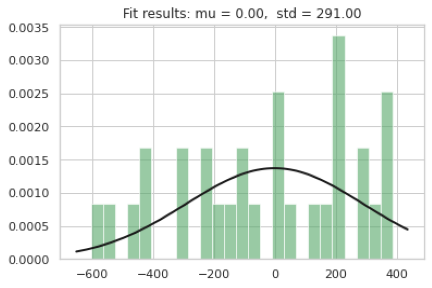}
  \includegraphics[scale=0.3]{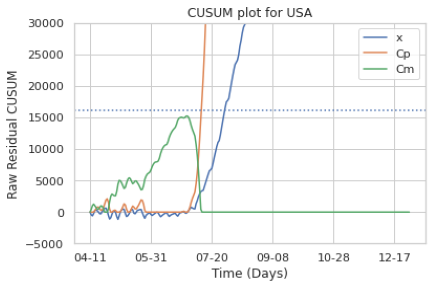}
  \includegraphics[scale=0.3]{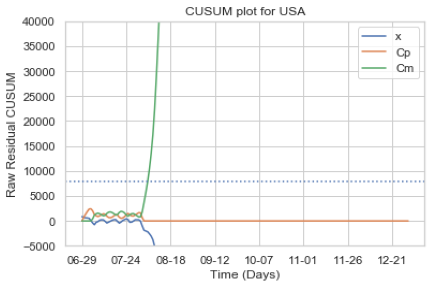}
  \includegraphics[scale=0.3]{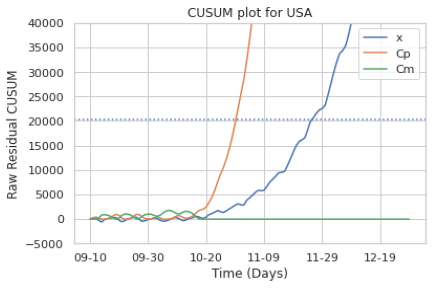}
  \caption{For US data, CUSUM algorithm gives 3 phase change detection points on 11 Jul, 05 Aug and 02 Nov 2020.}
    \label{Figure:US_CD}
\end{figure}

We included a constant factor $c$ using $h= c (5\, \sigma) $ to suit our case. Assuming normality of residuals, we fit a normal distribution on the training data residuals and estimate sample mean and sample standard deviation. Using these CUSUM parameter estimates we draw control charts.
Figure \ref{Figure:US_CD} shows the result of the algorithm for the whole period in our dataset for the US, where three change points are detected. Recall that by construction, the points detected are almost surely larger than the ``real'' change in phase, because there is a detection delay, and also because we use the data for the deaths: in all likelihood a change in phase will entail some delay before the death counts reflect a significant change in trend.

\begin{figure}[h!]
\centering
  \includegraphics[scale=0.45]{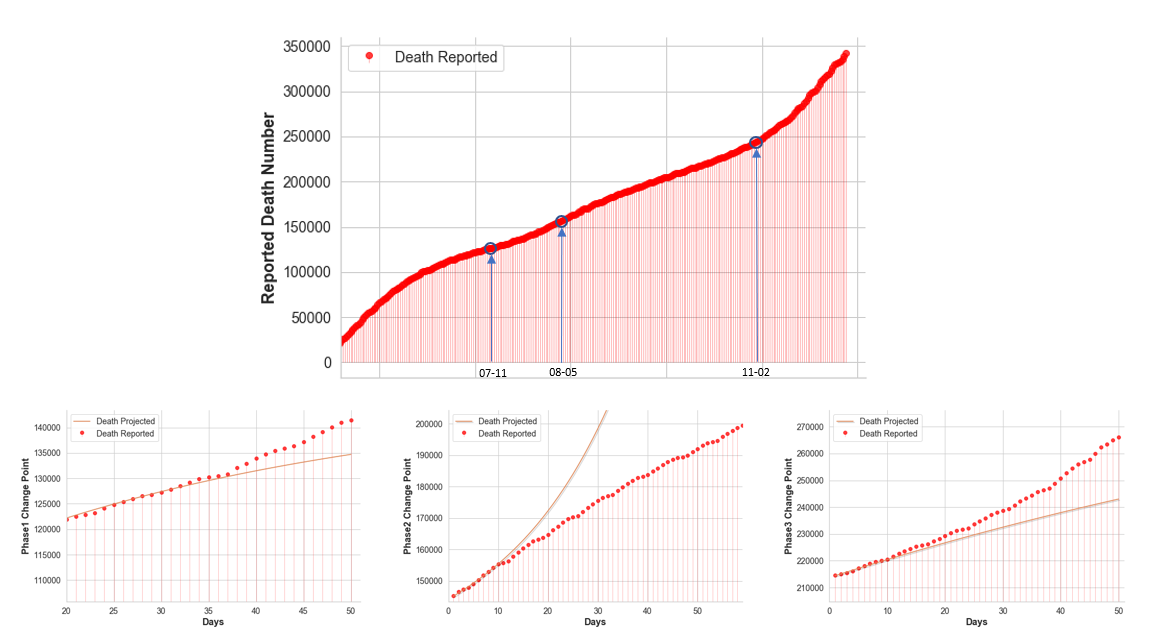}
  \caption{For US data, CUSUM algorithm gives 3 phase change detection points from 11 April 2020 to 29 Dec 2020: on 11 Jul, 05 Aug and 02 Nov, 2020.}
    \label{Figure:US_CDplots}
\end{figure}

Figure \ref{Figure:US_CDplots} provides a visualization of the data corresponding to the number of deaths reported for the US, three change points are marked in the plot. To interpret the results we have added a ``close-up'' around each change point showing what the previous phase would predict against the actual data. Looking at the plots, phase 2 presented a marked spike in the pandemic, compared to phase 1. In contrast, phase 3 has less growth than predicted under the regime from phase 2, but then phase 3 spikes again. The 11 July spike in number of deaths corresponds to a spike in the number of infected people occurring around Memorial Day. It is well known that this period was characteristic of public demonstrations, opening of restaurants and large public gatherings where people were not forced to wear masks and observe social distancing. Shortly afterwards the alarm was raised and measures were taken to impose stricter restrictions on social behavior, which explains our second change point around 05 August, where the spread of the virus seemed to be better controlled. The next change point shows a marked spike and occurs around 02 Nov, right after the Halloween celebrations, and we can see a small increase in slope (not detected by our algorithm) around the important Thanksgiving holiday. In spite of repeated warnings to the public, people traveled and kept plans for family reunions in this important holiday. It is highly possible that the spike detected by our algorithm was in part caused by this sudden change in behavior, where people were not vigilant of the preventive measures.

\begin{figure}[h!]
\centering
  \includegraphics[scale=0.35]{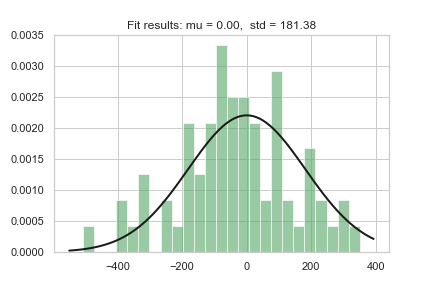}
  \includegraphics[scale=0.35]{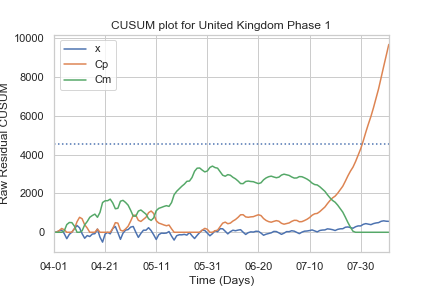}
  \includegraphics[scale=0.35]{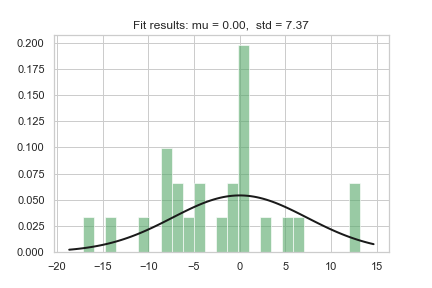}
  \includegraphics[scale=0.35]{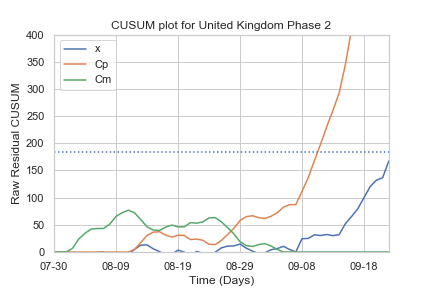}
 \caption{UK: CUSUM algorithm gives 2 phase change detection. The first one is detected on 31 July, the second on 11 September, 2020.}
    \label{Figure:UK_CD}
\end{figure}

Figure \ref{Figure:UK_CD} shows the results of the algorithm for the UK, where two change points are detected, one on 31 July and the other one on 11 September.

\begin{figure}[h!]
\centering
  \includegraphics[scale=0.45]{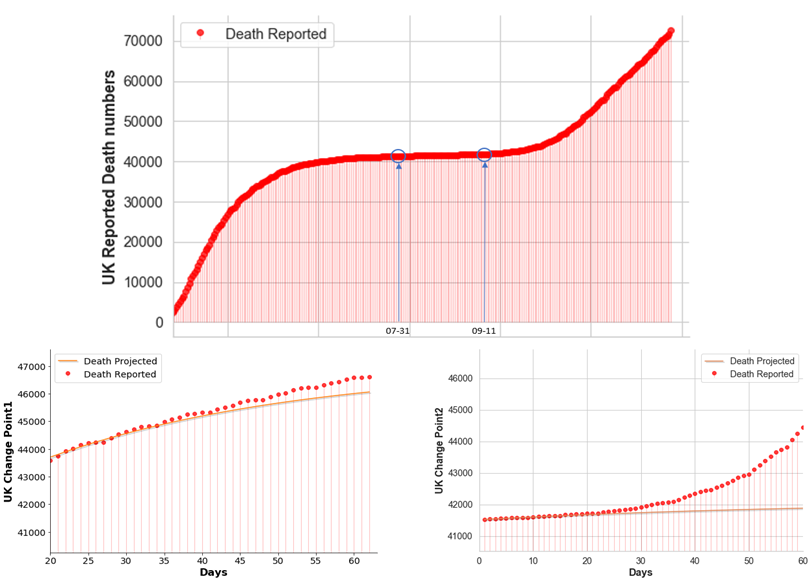}
  \caption{For UK data, CUSUM algorithm gives 2 phase change detection points from 01 April 2020 to 26 Dec 2020: on 31 Jul on 11 Sep, 2020}
    \label{Figure:UK_CDplots}
\end{figure}

To visualize the dynamics of the pandemic for the UK data, Figure \ref{Figure:UK_CDplots} shows the number of infected people, where we have marked the detected change points. It is worth noticing that both changes describe a worsening of the pandemic. In July, the UK government allowed no quarantine requirements for returning passengers from 59 countries. There were no restrictions on outdoor gatherings and people were not made to wear masks mandatorily. All these led to the occurrence of 2nd phase of infection which our model corroborates \citep{UK-times}. Table~\ref{table:2} shows the parameters of the model for US and UK on each of the phases detected.

\begin{table}[h!]
\centering
\begin{tabular}{l|llll|lll}
\hline
\hline
\textbf{Paras.} & US p1 & US p2 & US p3 & US p4 &UK p1 &UK p2 & UK p3 \\
\hline
$\beta$ & 1.0269 & \textbf{0.68} & 1.0673 & \textbf{1.493} & 1.485 & \textbf{0.907} & \textbf{0.089} \\
$c_1$ & 0.993 & \textbf{1.437} & \textbf{1.199} & \textbf{3.598} &0.704 & \textbf{1.556} & \textbf{1.741} \\
$c_2$ & 0.199 & \textbf{0.100} & 0.119 & 0.913 &0.124 & 0.115 & \textbf{5e-06} \\
%$\mathcal{R_0}$ & 4.029 & \textbf{2.21} & 4.099 & \textbf{6.193} &4.954 & \textbf{1.008} & 1.000 \\
$\sigma$ & 0.0321 & \textbf{0.05} & 0.0150 & 0.0150 &0.0517 & 0.0500 & \textbf{0.0100} \\
$M_{21}$ & 0.300 & \textbf{0.0001} & 0.347 & 0.897 &0.305 & \textbf{0.988} & 0.990 \\
$N_{21}$ & 0.989 & \textbf{0.0002} & 0.600 & 0.500 & 0.439 & \textbf{0.700} & \textbf{0.990} \\
$\mathcal{d}$ & 0.00483 & 0.00663 & \textbf{0.000284} & 0.000201 & 0.01604 & \textbf{0.000100} & 0.000100 \\
$\mathcal{r}$ & 0.250 & 0.300 & 0.260 & 0.241 & 0.283 & \textbf{0.899} & \textbf{0.0894} \\
$\alpha$ & 0.0899 & 0.0850 & 0.0890 & 0.0900 & 0.1999 & 0.1683 & \textbf{0.7739} \\
$v$   & 2.4567    & 2.420    & 2.417    & 3.121 &2.670&4.569& \textbf{6.90}\\
\hline
\hline
\end{tabular}
\caption{Model parameters of U.S. and U.K.}
\label{table:2}
\end{table}

Our results indicate that change points corresponding to increases in the infection rates are generally characterized by an increase value of the parameters $c_1, c_2$. This is an observation that we verified also with other datasets including Texas and New York. Of all the parameters, these seem to be significant. We can observe from Figure \ref{Figure:US_CDplots} that the second change point in US, which is between phase 2 and 3, is different from the other two, where the model over-estimated the number of deaths. This supports the idea that effective regulations were carried out after the explosion of the Summer wave of pandemic, and the estimated parameters shown in Table \ref{table:2} confirm this, as $c_1$ decreased significantly from phase 2 to 3. We noticed that the change between phase 2 and 3 in UK does not follow this pattern. Rather, the parameter that is significantly different is $\alpha$. The parameter $\beta$ being smaller may well be related to people following isolation practices thus diminishing the contact rate in the overall population, in spite of which we see a spike. The parameter $\alpha$ is related to the {\em infectivity rate} of the virus.

Random mutations of the virus occur constantly. In particular, around 900 mutations have already been described for the SARS-CoV2, most of them {\em punctual}, many of them either self-repair or result in a less potent version of the virus. However, in early December, mutation D614G was described in UK as having higher infectivity. More than 1000 people were then studied that had been infected with this mutation, but as far as we are aware no-one knows the exact time when this mutation was introduced in the world. Our data seems to indicate that it happened in September. The receptor-binding site of the spike molecule in this mutation has higher affinity, resulting in higher infectivity rate (our $\alpha$). According to our estimates, the infectivity parameter $\alpha$ is about seven times higher in UK phase 3 as it is in US and UK for previous phases.

\begin{figure}[h!]
\centering
  \includegraphics[scale=0.50]{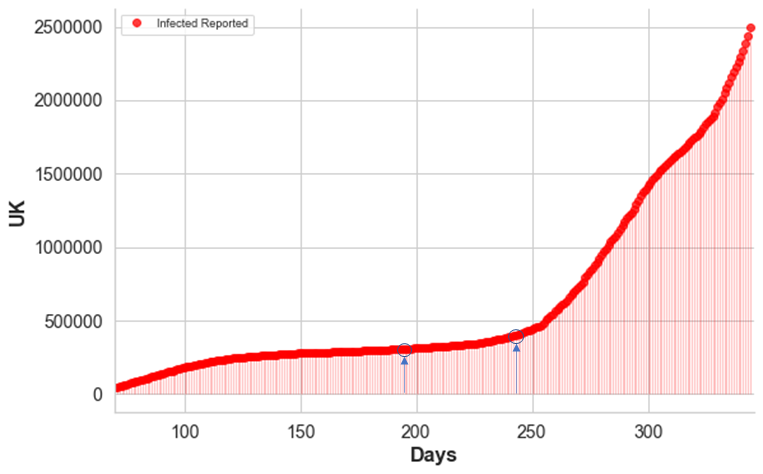}
  \caption{UK number of infected reported.}
    \label{Figure:UK-Infected}
\end{figure}

We finalize with an important observation. Because of the accuracy in the data, our change point detection is carried out on the residuals for the death counts (see \sect{stats} for details on dataset selection). However, what matters is an estimate of the actual change in phase in the spread of the virus. Figure~\ref{Figure:UK-Infected} shows the number of infections reported in UK. The two change points detected are marked in the plot.

To the naked eye, it is apparent that another change appears towards the end of the period. This coincides with the winter spike and it is perhaps related to the usual increase in activity of the corona viruses. The family of corona viruses is typically reactivated in winter, as is common--for example-- with the flu and the common cold, but this new variant of the virus seems to cause infections that are less fatal and with milder symptoms. It is possible that our algorithms did not detect the Winter spike because it works on the death count and up to today's date the sample size is not large enough.

\subsection{Two-way detection}
As mentioned earlier, by construction a change detection algorithm that builds the CUSUM processes $C_{p,m}(t)$ forward in time has a positive detection delay almost surely. In our case the detection helps to define the intervals of each of the phases in a pandemic. The goal of this research is to propose a methodology that can be used by social scientists, medical advisers and decision makers in order to better understand the spread of the virus and to have a model to analyze the impact of events or policies in the various phases of the pandemic under study.

With this in mind, it may be very important to obtain more accurate estimates of the true change points, which may provide invaluable information about the various factors that may have triggered the transitions between phases.

Let $T$ be the point detected by the algorithm, going forward in time as explained in the previous section. Two-way detection follows a simple principle. Look at the process in {\em reversed time}, starting from a point $\tau>T$ and going backwards in time. Given the parameter $\th$ that fits the model on the new phase (from $T$ to $\tau$), we now need to solve the equations in order to find the number of deaths going backwards, assuming $\th$ governs the dynamics of the process. Once this  is calculated, the residuals can be evaluated using the data, and change detection can be carried out in exactly the same way as before, but going backwards in time.

Our SEIDR model is a system of autonomous ordinary differential equations in dimension $9$. Call $x(t)$ the vector-valued process, and denote the system of equations by
\[
\frac{d x(t)}{dt} = f(\th; x(t)); \quad x(0)=x_0.
\]
Define now the process $y(t) = x(\tau-t)$. Our hypothesis will now test for detection on the process $y(t)$ to discover when the (new) parameters $\th$ no longer describe the (past) observed data. Simple change of variables imply that
\[
\frac{d y(t)}{dt} = -f(\th; x(\tau-t)) = -f(\th;y(t)), \quad y_0 = x(\tau).
\]

Thus, using the reversed time ODE, we can perform detection, providing now a new estimate $T'$ of the time when the new phase no longer describes the past.

The final estimate for the change point is declared to be $(T+T')/2$: the mid point between the two detected points. We should remark that some packages have a built in ODE solver, however the most robust method if we wish to use the two-way detection is the {\em leap frog} method for solving the ODEs, because this method is time reversible, and thus more consistent when considering the evolution of the model forwards and backwards in time.

\section{Data-driven Phase Detection}
\label{sec:dpd}
 % to make into a new file?
%{\color{red} Yuansan to change to automatic labeling and referencing}.
In \sect{stats} we explained how validation and parameter estimation are carried out for the SEIDR model using past data. In \sect{cpd} we explained how data can be used to detect changes in the phases of the model.

This section presents our novel methodology that allows for concurrent estimation and phase detection for data streaming. While the previous sections used data in offline mode, the algorithms here can be implemented online to better inform the experts about the evolution of the pandemic in their socio-political regions of interest. As the spread of the virus evolves, our algorithm estimates both the parameters of the SEIDR model as well as phase change detection.

Our model works with a sliding window of $\Delta$ days. From our extensive testing on the COVID-19 data, we propose $\Delta=30$. Initially, the data from $\Delta$ days is used to fit and validate the model, as done in \sect{stats}. Call $\tau_l=0$, $\tau_r = \Delta$, and the first phase is defined as $[\tau_l, \tau_r]$. The time $\tau_r$ is the time when dynamic estimation starts, as follows.

{\bfseries Dynamic change detection:} Using the parameter $\th$ calculated for the current phase using data before time $\tau_r$  compute $X_t, C_p(t), C_m(t)$ at time $t$ and compare to threshold $h$. This process is iterative and updates these quantities online, streaming data for $t> \tau_r$. There are two possible scenarios:

\begin{itemize}
\item No change is detected for $\Delta $ more days (that is, for days in the interval $\tau_r < t \le \tau_r+\Delta$). If this is the case, fit the model parameters $\th$ again using the data from $[\tau_l, \tau_r+\Delta]$. Set $\tau_r:=\tau_r+\Delta$ and continue with the dynamic change detection.
\item Change is detected at time $T< \tau_r+\Delta$. Complete gathering data for the period $[T, T+\Delta]$. Set $\tau_l=T, \tau_r=\tau_l+\Delta$ and declare the new phase on $[\tau_l, \tau_r]$. Fit model parameters for the new phase with this data. From $\tau_r$ onwards perform dynamic change detection as the data becomes available daily.
\end{itemize}

\begin{algorithm}
\caption{Dynamic Phase Detection}
\begin{algorithmic}[1]

\Procedure{DynamicPh}{$\theta(\tau_l,\tau_r)$,$X_t$, $C_p$[t-1], $C_m$[t-1]}
    \State $daysCounter$ $\gets$ 0
    \While{(t $>$ $\tau_r$)}
        \State   $C_p$[t] $\gets$ max(0,$C_p$[t-1] + $X_{t}$ - ($\mu_0$ + $K$))    \Comment{Get Cusum vals}
        \State   $C_m$[t] $\gets$ max(0,$C_m$[t-1] - $X_{t}$ + ($\mu_0$ - $K$))
        \State   $daysCounter$ $\gets$ $daysCounter$ + 1
        \If{($C_p$[t] $<$ $h$ $\&$ $C_m$[t] $<$ $h$)}
            \If{($daysCounter$ $>$ $\Delta$)}
                \State $\tau_r$ $\gets$ $\tau_r$ + $\Delta$
                \State $\theta$ $\gets$ $modefit(\tau_l,\tau_r)$
                \State $daysCounter$ $\gets$ 0
            \EndIf
        \Else
            \State     $\tau_l$ $\gets$ t \Comment{New Phase Detected !!}
            \State $\tau_r$ $\gets$ t + $Delta$
            \State $\theta$ $\gets$ $modelfit(\tau_l,\tau_r)$
            \State $daysCounter$ $\gets$ 0
        \EndIf
    \EndWhile  \label{DyCDPseudo}
\EndProcedure

\end{algorithmic}
\end{algorithm}

We conclude this section showing the results of robust predictions, where our model is used to predict the near future evolution of the pandemic using different values of the parameter $c_1$. According to our analysis, it is apparent that this is the parameter that has a greater influence in the sudden change when the infection rate blows up.

\begin{figure}[h!]
  \centering
  \includegraphics[scale=0.5]{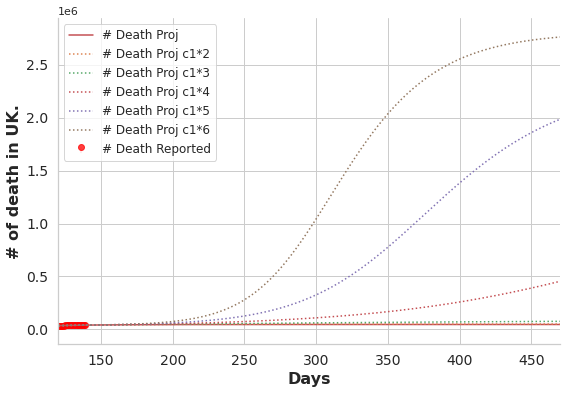}
  \caption{Model predictions for UK under various scenarios.}
  \label{Figure:UK-robust}
\end{figure}

Figure \ref{Figure:UK-robust} shows such scenarios. Using the current estimate $c_1$ the plots show the predictions if this parameter were to increase $k$ times, for $k=1,\ldots,6$. The larger value $6 c_1$ leads to a staggering increase of deaths in a short period of time, reaching over 280,00 in 100 days, while the value $4 c_1$ yields 70,000 in the same period of time. The projected number using $c_1$, however, is stable. This kind of analysis can be put to practice in order to evaluate the possible consequences of relaxing social restrictions in a population. How to actually achieve a particular value for $c_1$ is outside the scope of the present paper, but we plan to pursue research in that direction in the near future.

%To the left we show three scenarios for about a year. To the left we show an extreme case, where we can compare the results if $c_1$ were to be 10 times smaller (worse scenario), which leads to a staggering increase of deaths in a short period of time, reaching over $150,000$ under a year. In contrast, were $c_1$ multiplied by 4 then the number of deaths would not greatly surpass the initial $45,000$. The plot to the right considers

\section{Concluding Remarks}
We have shown that our proposed 2-group SEIDR model with phase change detection simulates the spread of COVID-19 disease well. Acknowledging, the spatiotemporal disease dynamics, we have proposed a methodology that uses the reported data to draw its conclusions and does not rely on policy or decision-makers. It is rather a tool to assists such institutions for better analysis and disease phase length estimations. Another possible use of our model is to compare different geo-political regions based on their infection reporting accuracy and data quality by using our vertical adjustment.

Each parameter in our modeling relates to either public or government or disease behavior. Exploiting these parameters, policymakers will be able to better understand their impact and predict future trends. This will help the region make more informed decisions on lockdown strategies.

Our proposed combination of statistical fitting with phase change detection can be used by other researchers to study past pandemics and its evolution. This will not only help us understand the behavior of people but also educate us for future potential pandemics. Our forward and backward CUSUM methodology will be particularly very useful in studying past events to analyze if the policies used were congruent with the reported data.

\bibliographystyle{apacite}
\bibliography{acl2020.bib}

\end{document}